\documentclass[a4paper,11pt]{article}
\usepackage{jheppub} 

\def\be{\begin{equation}}
\def\ee{\end{equation}}
\def\bea{\begin{eqnarray}}
\def\eea{\end{eqnarray}}
\def\nnb{\nonumber}

\newcommand{\Lu}{\mathcal{L}}
\newcommand{\f}{\frac}
\newcommand{\scs}{\scriptscriptstyle}

\title{\boldmath Effective Field Theories in $R_\xi$ gauges }

\author[a,1]{M. Misiak,\note{Corresponding author.}}
\author[a,b]{M. Paraskevas,}
\author[a]{J. Rosiek,}
\author[b]{K. Suxho}
\author[a]{and B. Zglinicki}

\affiliation[a]{Institute of Theoretical Physics, Faculty of Physics,University of Warsaw,\\ 
                Pasteura 5, PL 02-093, Warsaw, Poland}
\affiliation[b]{Department of Physics, University of Ioannina, GR 45110, Ioannina, Greece}

\abstract{In effective quantum field theories, higher dimensional operators
  can affect the canonical normalization of kinetic terms at tree level. These
  contributions for scalars and gauge bosons should be carefully included in
  the gauge fixing procedure, in order to end up with a convenient set of
  Feynman rules. We develop such a setup for the linear $R_\xi$-gauges. It
  involves a suitable reduction of the operator basis, a generalized gauge
  fixing term, and a corresponding ghost sector. Our approach extends previous
  results for the dimension-six Standard Model Effective Field Theory to a
  generic class of effective theories with operators of arbitrary dimension.}

\begin{document} 
\maketitle

\section{Introduction} \label{sec:intro}

The Higgs boson of the Standard Model (SM) remains the only particle
discovered at the LHC so far, despite several years of searches at
$13\,$TeV~\cite{Tanabashi:2018oca}. Thus, it becomes more and more likely that
a sizeable energy gap between the new physics scale and the electroweak scale
is present. In this region, the most convenient calculational framework is an
Effective Field Theory (EFT) with only the SM degrees of freedom, the
so-called SMEFT~\cite{Buchmuller:1985jz, Grzadkowski:2010es, Henning:2015alf}.
Higher-dimensional SMEFT operators can account for the neutrino masses and
mixings, as well as for other indirect signals for beyond-SM physics that
emerge with growing statistical significance in the magnetic moments of
leptons~\cite{Tanabashi:2018oca, Parker:2018xxx}, and in several $B$-meson
decay channels~\cite{Aebischer:2018iyb}.

Practical calculations within the SMEFT require introducing convenient
gauge-fixing terms. In particular, it has been observed in
Refs.~\cite{Dedes:2017zog, Helset:2018fgq} that effects of higher-dimensional
operators should be taken into account in the definition of $R_\xi$ gauges to
remove tree-level mixing between the gauge and would-be Goldstone (WBG)
bosons, and to preserve simple relations among various masses. Explicit
expressions for the dimension-six operator effects have been derived.

In the present paper, we extend the analysis of Refs.~\cite{Dedes:2017zog,
Helset:2018fgq} to a wide class of EFTs with operators of arbitrary
dimension. We consider a generic local EFT with linearly realized gauge
symmetry, and matter fields in arbitrary representations of the gauge group
$G$. The matter fields are assumed to contain spin-0 fields that develop
Vacuum Expectation Values (VEVs), and the Higgs mechanism takes place, giving
mass to some of the gauge bosons. In such a case, tree-level mixing between
the gauge and WBG bosons arises not only from the dimension-four part of the
Lagrangian, but also from operators of arbitrarily high dimension. To remove
such a mixing with the help of $R_\xi$ gauge-fixing terms, we are going to
arrange our operator basis in a particular manner, using the Equations of
Motion (EOM) to simplify the bilinear terms. Next, after introducing the
$R_\xi$ gauge fixing in an appropriate manner, we shall verify that the
standard relations between the gauge and WBG boson masses remain
valid. Explicit expressions for the ghost terms and BRST transformations
will be given.

Our article is organized as follows. In the next section, the operator basis
simplification with the help of EOM is described. Section~\ref{sec:gf} is
devoted to defining the gauge fixing and deriving the mass relations. In
Section~\ref{sec:gs}, the ghost sector and the BRST transformations are
specified. We conclude in Section~\ref{sec:concl}. In Appendix~\ref{app:gi},
we recall the arguments behind constructing the EFT operators from products of
fields and their covariant derivatives. Appendix~\ref{app:xi} is devoted to
generalizing our results to the case of several distinct gauge-fixing
parameters. Appendix~\ref{app:re} summarizes basic expressions for complex
scalar field representations of $G$ in the real notation. The specific
example of SMEFT is discussed in Appendix~\ref{app:smeft}.

\section{Operator basis reduction} \label{sec:or}

Let us consider an EFT that arises after decoupling~\cite{Appelquist:1974tg}
of heavy particles whose masses are of the order of some scale $\Lambda$. We
assume that the original theory at that scale is perturbative. The resulting
EFT describes dynamics of light fields (with masses $\ll \Lambda$) at energy
scales much lower than $\Lambda$. It is given by the following Lagrangian
\be
\label{lagr}
\Lu ~=~ \Lu^{(4)} ~+~ \sum_{k=1}^\infty \f{1}{\Lambda^k} \sum_i C_i^{(k+4)} Q_i^{(k+4)}\; ,
\ee
where $\Lu^{(4)}$ is the dimension-four (``renormalizable'') part of $\Lu$,
while $Q_i^{(k+4)}$ stand for dimension-$(k+4)$ local operators built out of
fields and their derivatives. They come with the Wilson coefficients
$C_i^{(k+4)}$. Throughout the paper, we work at an arbitrary but {\em fixed}
order $N$ in the $1/\Lambda^k$-expansion, i.e.\ we are going to neglect terms
of order $1/\Lambda^{N+1}$ or higher.

Spin-0 degrees of freedom can always be described in terms of real scalar
fields. They will be denoted collectively by $\Phi$.  We are interested in
situations when $\Phi$ acquires a non-vanishing VEV $\langle \Phi \rangle = v$
such that $\left|v\right| \ll \Lambda$. If $v$ is not a singlet under $G$,
some of the gauge fields $A_\mu^a$ become massive via the Higgs mechanism. We
assume absence of any other relevant contributions to the gauge boson masses
(like, e.g., contributions from fermion-antifermion pair condensation).

To keep the notation compact, we absorb the gauge couplings into the structure
constants $f^{abc}$ and the gauge group generators $T^a$. Then the field
strength tensor for all the gauge fields
\be
F^a_{\mu\nu} = \partial_\mu A^a_\nu - \partial_\nu A^a_\mu - f^{abc} A_\mu^b A_\nu^c\; ,
\ee
as well as the covariant derivatives of $\Phi$ and $F_{\mu\nu}$
\be
D_\mu \Phi = (\partial_\mu + i A^a_\mu T^a) \Phi, \hspace{2cm} 
(D_\rho F_{\mu\nu})^a = \partial_\rho F^a_{\mu\nu} - f^{abc} A^b_\rho F^c_{\mu\nu}
\ee
are given by the above short-hand expressions even if the gauge group G is not
simple, and/or $\Phi$ resides in a reducible representation. The generators
$T^a$ of this representation are both hermitian and antisymmetric, which means
that all their components are imaginary (see Appendix~\ref{app:re}).

Our goal in this section is selecting and simplifying all the terms in
$\Lu$~\eqref{lagr} that matter for tree-level two-point Green's functions for
the scalar and gauge fields. Thus, we shall consider such operators $\Lu$ that
contain bilinear terms in $\varphi = \Phi - v$ and in the gauge fields
$A^a_\mu$.  Before gauge fixing, the part of $\Lu^{(4)}$ that matters for our
considerations is the one containing solely $\Phi$ and $A_\mu^a$
\be \label{lagr4PhiA}
\Lu^{(4)}_{\Phi,A} = \f12 (D_\mu \Phi)^T (D^\mu \Phi) -\f14 F^a_{\mu\nu} F^{a\,\mu\nu} - V(\Phi)\; ,
\ee
where $V$ is the scalar potential. Fermionic matter fields with half-integer
spins have no effect on the bilinear terms we are after. We assume that no
other bosonic degrees of freedom but $\Phi$ and $A_\mu^a$ are present in our
EFT.\footnote{
We neglect the gravitational interactions.}
Thus, all the terms in $\Lu$~\eqref{lagr} that contain any other field but
$\Phi$ or $A_\mu^a$ are going to be ignored from now on.

In this section, we treat $\Lu$~\eqref{lagr} as the tree-level Lagrangian,
before introducing gauge fixing and/or ultraviolet counterterms. It means that
$\Lu$ is a linear combination of gauge-invariant operators that are built of
gauge field strength tensors, matter fields, and their (multiple) covariant
derivatives (see Appendix~\ref{app:gi}). It can be simplified with the help of
EOM that take the following form
\be \label{eom}
D^\mu D_\mu \Phi = \boxed{HL}\; , \hspace{2cm} (D^\mu F_{\mu\nu})^a = \boxed{HL}\; ,
\ee
where $\boxed{HL}$ stands for either higher-dimensional or lower-derivative
terms. By lower-derivative terms we mean terms containing a lower number of
covariant derivatives.  If two terms contain the same number of covariant
derivatives, then the one containing a lower number of field strength tensors
is considered to be ``lower''.

Simplification with the help of EOM may be understood as writing as many
interactions as possible in terms of expressions that vanish by EOM, i.e.\
EOM-vanishing operators. Green's functions with single insertions of such
operators have no effect on on-shell amplitudes~\cite{Politzer:1980me,
KlubergStern:1975hc, GrosseKnetter:1993td, Arzt:1993gz, Simma:1993ky,
Wudka:1994ny, Manohar:2018aog, Criado:2018sdb}. For this reason, such
operators are often considered redundant, and are removed from the operator
basis. Whenever multiple insertions of them matter, the right tool for the
operator basis simplification are certain field redefinitions rather than the
EOM themselves. However, the ultimate effect of such a procedure is that the
simplified basis contains no EOM-vanishing operators (see, e.g., section 5 of
Ref.~\cite{Criado:2018sdb}).

One should not forget that EOM-vanishing operators may be relevant as
ultraviolet counterterms in renormalizing off-shell Green's functions, along
with certain gauge-variant operators~\cite{KlubergStern:1975hc}.  Since our
current discussion is restricted to the tree-level Lagrangian only, we can
determine its structure by setting all the EOM-vanishing and gauge-variant
terms to zero.

Any gauge-invariant operator containing $n$ scalar fields, $m$ field
strength tensors, and $k$ covariant derivatives will be symbolically
denoted by $\Phi^n F^m D^k$. The power $k$ must be even due to Lorentz
invariance.  We shall show that the EOM allow to bring the Lagrangian into
such a form that only $\Phi^n$, $\Phi^n D^2$ and $\Phi^n F^2$ operators matter
for the scalar and gauge boson bilinear terms.

First, let us realize that the bilinear terms in question can only originate
from such products where at most two objects have vanishing VEVs. By
``objects'' we mean tensors $F$, scalar fields $\Phi$ or (multiple) covariant
derivatives of them written in any ordering.  For instance, a contraction of
four objects like $(D_\mu \Phi)^T (D^\mu \Phi) F^a_{\nu\rho} F^{a\,\nu\rho}$
does not affect the bilinear terms because all the four objects have vanishing
VEVs. On the other hand, $(\Phi^T \Phi) (\Phi^T D^\mu D^\nu D_\mu D_\nu \Phi)$
can potentially affect the bilinear terms, because only one object there (the
fourth covariant derivative of $\Phi$) has a vanishing VEV.

We are not going to consider contractions involving
$\varepsilon_{\mu\nu\alpha\beta}$ in a separate manner. Since all our spin
$\neq 0$ objects (tensors F and covariant derivatives of something) have
vanishing VEVs, the $\varepsilon$ tensor can contract no more than two objects
in the operators of interest. In such a case, it is easy to convince oneself
that any contraction with $\varepsilon$ can be written in terms of the dual
tensor $\widetilde{F}$ and no explicit $\varepsilon$. In our considerations
below, whenever $F$ is being mentioned, in may sometimes mean also
$\widetilde{F}$. If $F$ is inserted into the l.h.s. of its EOM~\eqref{eom},
doing the same for $\widetilde{F}$ means using the Bianchi identity $(D^\mu
\widetilde{F}_{\mu\nu})^a = 0$.

We shall proceed as follows. After picking up an operator of dimension $d$
that may matter for the bilinear terms in $\varphi$ or $A^a_\mu$, we are going
to use the EOM to express it in terms (``reduce it to'') either dimension $>
d$ operators or lower-derivative ones. Such a procedure can be applied
subsequently starting from the operators of lowest dimension, and then
proceeding to higher dimensions. At a given dimension, we start from operators
with the highest number of derivatives, and then proceed to lower-derivative
ones. In this way, at each given dimension, all the operators with derivatives
of $F$, and all the operators with multiple derivatives of $\Phi$ will turn
out to be reducible. Higher-dimensional terms arising from the EOM at each
step will eventually get shifted beyond the order $1/\Lambda^N$ that has been
assumed to be the highest one in our treatment of the EFT.

Let us begin with considering an operator containing $D_{\mu_1} \ldots
D_{\mu_k} \Phi$ where some of the Lorentz indices are contracted. We can
permute the derivatives (at the cost of introducing lower-derivative terms via
$[D_\mu,D_\nu] \sim F_{\mu\nu}$) to express the considered $k$-th derivative
in terms of objects containing either more $F$ tensors or $D_\mu D^\mu
\Phi$. In the latter term, we apply the EOM for $\Phi$ to reduce it to either
higher-dimensional or lower-derivative terms. This way we can eliminate all
the terms with contracted derivatives of a single $\Phi$, up to the highest
dimension we would like to include.

Next, we consider a Lorentz-scalar operator containing $D_{\mu_1} \ldots
D_{\mu_k} \Phi$ with $k\geq2$, where none of the indices are contracted among
themselves. They must be contracted with other objects carrying Lorentz
indices to give an invariant operator. Since we allow only for two objects
with non-vanishing VEVs, this other object must be either $D^{\mu_{\sigma(1)}}
\ldots D^{\mu_{\sigma(k)}} \Phi$, where $\sigma$ is some permutation, or a
(multiple) derivative of $F$.\footnote{
For $k=2$, we skip the option with $F^{\mu_1\mu_2}$. In such a case,
$D_{\mu_1} D_{\mu_2} \Phi$ can be replaced by $[D_{\mu_1}, D_{\mu_2}] \Phi
\sim F^a_{\mu_1\mu_2} T^a \Phi$, and we obtain an operator with double $F$ and
no derivatives of $\Phi$, to be considered below.}
Shifting one of the $D^{\mu_{\sigma(i)}}$ derivatives ``by parts'', we obtain
either contracted derivatives acting on a single $\Phi$ (discussed above), or
operators with three terms whose VEVs vanish.

Thus, what we have shown so far is that all the operators containing second
and higher derivatives of $\Phi$ can be either removed via EOM or do not
affect the bilinear terms.

Now, let us consider operators containing at least one gauge field strength
tensor $F$. These tensors have vanishing VEVs, so only one or two such tensors
are allowed in the operators that affect the bilinear terms. If there is only
one such tensor, its indices can be contracted either with some of the
derivatives acting on the very $F$, or with the first derivative of
$\Phi$. Since only a single derivative of $\Phi$ is allowed at this point, one
of the indices of $F$ must be contracted with one of the derivatives acting on
the very $F$. After a permutation of derivatives (which brings to life more
$F$'s), we find a contraction $[(...)  D^\mu F_{\mu\nu}]^a$ that reduces via
EOM to higher-dimension or lower-derivative operators. Thus, all the operators
with single $F$ can either be removed via EOM or do not affect the bilinear
terms.

It remains to discuss operators with double $F$. In this case, no derivative
of $\Phi$ allowed because three objects would have vanishing VEVs. Let us show
that operators with derivatives acting on $F$ can be removed, too. None of the
$F$'s can be fully contracted with derivatives acting on any single object
because this would bring us to the case with at least three $F$'s via~
$F^{\mu\nu} D_\mu D_\nu = \f12 F^{\mu\nu} [D_\mu, D_\nu] \sim F^{\mu\nu}
F_{\mu\nu}$.  On the other hand, if $F$ is contracted with at least one
derivative acting on the very $F$, we proceed as in the previously discussed
case with a single $F$, arrive at the EOM for $F$, and get moved to
higher-dimension or lower-derivative operator classes. Thus, the only
remaining options for contractions of Lorentz indices are:
\be
Y^{ab}~[(\ldots) (D_\mu F_{\nu\rho})]^a [(\ldots) (D^\mu F^{\nu\rho})]^b
\hspace{1cm} \mbox{or} \hspace{1cm}
Y^{ab}~[(\ldots) (D_\mu F_{\nu\rho})]^a [(\ldots) (D^\nu F^{\mu\rho})]^b\; ,
\ee
where $(\ldots)$ stand for possible extra derivatives, while $Y^{ab}$ is built
out of the $\Phi$ fields only (with no derivatives). The first of the above
options can be converted to the second one with the help of the Bianchi
identity\footnote{
If the considered tensor is $\widetilde{F}$, we do the same via the EOM for
$F$, up to higher-order and/or lower-derivative terms.}
$(D_{[\mu} F_{\nu\rho]})^a=0$. In the second option, we shift $D_\mu$ from the
middle term ``by parts''. After doing this, we ignore all the terms with
derivatives of $\Phi$ or commutators of covariant derivatives because they
contain more than two terms with vanishing VEV's. This way we arrive at
\be
Y^{ab}~[(\ldots) (F_{\nu\rho})]^a [(\ldots) (D^\nu D_\mu F^{\mu\rho})]^b\; ,
\ee
where the EOM for $F$ can be applied, reducing the considered expression to
higher-dimensional or lower-derivative terms.

At this point, our EFT Lagrangian (at the considered arbitrary but fixed order
in $1/\Lambda$) already has the desired property, namely that only the
$\Phi^n$, $\Phi^n D^2$ and $\Phi^n F^2$ operators matter for the scalar and
gauge boson bilinear terms. As far as the $\Phi^n F^2$ operators are
concerned, we can now remove the possibility $Y^{ab} F^a_{\mu\nu}
\widetilde{F}^{b\,\mu\nu}$, in which case the only possible bilinear term
is a total derivative.

\section{Gauge fixing} \label{sec:gf}

We have organized our Lagrangian in such a way that only operators with at
most first derivatives of $\Phi$ and no derivatives of $F$ matter for the
bilinear terms in $\varphi = \Phi - v$ and $A^a_\mu$. All such operators
belonging to the classes $\Phi^n D^2$ and $\Phi^n F^2$ form a
(gauge-invariant) part of the Lagrangian $\Lu$~\eqref{lagr} that can be
written in the following form
\be \label{Lu}
\Lu_{J,K} = -\f14 F^a_{\mu\nu}\, J^{ab}[\Phi]\, F^{b\,\mu\nu} ~+~ 
\f12 (D_\mu \Phi)_i\, K_{ij}[\Phi]\, (D^\mu \Phi)_j\; , 
\ee
where the $\Phi$-dependent matrices $J$ and $K$ are symmetric. They form a
series in $1/\Lambda$ with the leading ($1/\Lambda^0$) contributions coming
from $\Lu^{(4)}_{\Phi,A}$~\eqref{lagr4PhiA}, and being equal to $\delta_{ij}$
and $\delta^{ab}$, respectively. Non-leading terms are polynomial in
$\Phi/\Lambda$, and depend on the Wilson coefficients.

The form of Eq.~\eqref{Lu} has been used in Ref.~\cite{Helset:2018fgq} to fix
the gauge for the dimension-six SMEFT using the Background Field Method
(BFM). In that paper, one can find explicit expressions for $J$ and $K$ at
${\mathcal O}(1/\Lambda^2)$, as they appear in this particular EFT. Note
that the operator reduction presented in the previous section ensures that
Eq.~\eqref{Lu} still holds at higher orders in the SMEFT expansion. This is
required for extending the BFM beyond the dimension-six level, and it would
not be guaranteed without the prior use of EOM. If the higher-derivative terms
were not eliminated from the bilinear terms via the EOM, they would need to be
treated as interactions affecting two-point functions at the tree level.
 
Let us note that a derivation of explicit expressions for $J[\Phi]$ and
$K[\Phi]$ in cases when multiple insertions of EOM-vanishing operators might
matter should be based on field redefinitions~\cite{Manohar:2018aog,
Criado:2018sdb} rather than simply setting the EOM-vanishing operators to
zero.

We now return to our main task, which is to present a formalism for $R_\xi$
gauges in generic EFTs. We focus on the bilinear terms in Eq.~\eqref{Lu} that
arise when $J$ and $K$ are set to their expectation values, i.e.
\be \label{eq:gdefc}
J^{ab}[\Phi] \to J^{ab}[v]\equiv J^{ab}
    \hspace{1cm} \mbox{and} \hspace{1cm}
    K_{ij}[\Phi] \to K_{ij}[v]\equiv K_{ij}\; .
\ee
Now $\Lu_{J,K}$ can be written as
\be \label{L}
\Lu_{J,K} = -\f14 A_{\mu\nu}^T \,J \,A^{\mu\nu} ~+~ \f12 (D_\mu \Phi)^T \, K\, (D^\mu \Phi) + \ldots\; ,   
\ee
where $A_{\mu\nu}^a \equiv \partial_\mu A^a_\nu - \partial_\nu A^a_\mu$.  The
terms denoted with ellipses in the above equation describe interactions of
three or more fields, and are irrelevant for our $R_\xi$ gauge fixing
procedure. In specific models, the structure of the matrices $J$ and $K$
can be constrained by the remaining local or global symmetries of the theory. This
happens, in particular, in the SMEFT -- see Appendix D.

Expanding the covariant derivatives in the second term of Eq.~\eqref{L}, using
integration by parts, and taking into account that $K$ is a symmetric matrix,
one obtains the usual ``unwanted'' term
\be \label{LAphi}
\Lu_{A\varphi} = -i\left( \partial^\mu A_\mu^a\right) \left[ \varphi^T K T^a v \right]\; ,
\ee
that describes the gauge and WBG boson mixing. In a convenient setup for
perturbative calculations, the $R_\xi$ gauge fixing term should remove this
unwanted mixing.

The scalar fields in the square brackets in Eq.~\eqref{LAphi} are identified
as the WBG bosons. They correspond to excitations of the scalar fields
$\varphi$ along orbits of the gauge group in directions of the broken
generators. The remaining excitations of $\varphi$ (that correspond to
physical scalars) span a space that is also determined by Eq.~\eqref{LAphi}.
It is the space that is orthogonal to all the $T^a v$ vectors, with
``orthogonality'' defined by the scalar product $K$. Thus, operators
suppressed by powers of $\Lambda$ that affect $K$ do have influence on our
identification of the WBG and physical excitations of $\varphi$.

Before introducing the gauge fixing, the WBG excitations are massless. It is
guaranteed by gauge invariance of the full scalar potential that includes both
$V(\Phi)$ from $\Lu^{(4)}$~\eqref{lagr4PhiA}, and all the relevant
contributions from higher-dimensional operators.

Let us now introduce the $R_\xi$ gauge fixing term
\be \label{Lgf}
\Lu_{GF} = -\f{1}{2\xi}\, {\mathcal G}^a J^{ab} {\mathcal G}^b
%
\hspace{1cm} \mbox{with} \hspace{1cm} 
%
{\mathcal G}^a = \partial^\mu A_\mu^a -i\xi  (J^{-1})^{ac} \left[  \varphi^T K T^c v \right]\; .
\ee
It is straightforward to check that the ``unwanted'' mixing of
Eq.~\eqref{LAphi} cancels in the sum $\Lu_{J,K} + \Lu_{GF}$. The bilinear
terms in this sum read
\bea
\Lu_{\rm kin,mass} &=& -\f14 A^T_{\mu\nu} J A^{\mu\nu} 
+\f12 A_\mu^a\left[ v^T T^a K T^b v \right] A^{b\,\mu}
+\f12 (\partial_\mu \varphi)^T K (\partial^\mu \varphi)\nnb\\
&& -\f{1}{2\xi}(\partial^\mu A_\mu)^T J (\partial^\nu A_\nu) -
       \f{\xi}{2} \left[ \varphi^T K T^a v
         \right](J^{-1})^{ab} \left[ v^T T^b K \varphi \right]\; . \label{psgmass}
\eea
The last term is the WBG boson mass matrix that comes solely from
$\Lu_{GF}$. The physical scalar mass terms (coming from the full scalar
potential) are not included in the above equation.

Let's diagonalize the above kinetic and mass terms. The matrices $J$ and
$K$ are symmetric and strictly positive-definite because $\left|v\right| \ll
\Lambda$.  Thus, they are diagonalizable and invertible. Moreover, they
possess positive-definite square roots that are also symmetric and invertible.
We can use them to redefine the scalar and gauge boson fields as follows:
\be \label{redef}
\tilde{\varphi}_i  = (K^\f12)_{ij} \varphi_j, \hspace{2cm}
\tilde{A}^a_\mu = (J^\f12)^{ab} A^b_\mu\; . 
\ee
After such a redefinition, we get
\bea
\Lu_{\rm kin,mass} &=& -\f14 \tilde{A}^T_{\mu\nu} \tilde{A}^{\mu\nu} 
                  +\f12 \tilde{A}_\mu^T (M^T M) \tilde{A}^{\mu}
                  +\f12 (\partial_\mu \tilde{\varphi})^T (\partial^\mu \tilde{\varphi})\nnb\\
&&                -\f{1}{2\xi}(\partial^\mu \tilde{A}_\mu)^T (\partial^\nu \tilde{A}_\nu) 
                  -\f{\xi}{2} \tilde{\varphi}^T (M M^T) \tilde{\varphi}\; . \label{mmt}
\eea
The kinetic terms have already acquired the canonical form, while the mass
matrices are given in terms of
\be \label{eq:mmatdef}
M_j^{~b} \equiv [K^\f12 (iT^a) v]_j\, (J^{-\f12})^{ab}\; .
\ee
The above matrix is not a square one (in general) because the scalars and
gauge bosons usually reside in representations of different
dimensionality. Below, we denote the number of real scalar fields by $m$, and
the number of gauge bosons by $n$, which means that $M$ is a real $m\times n$
matrix.

To diagonalize the mass matrices, one can apply the Singular Value Decomposition (SVD)
\be \label{svd}
M = U^T \Sigma V
\ee
with certain orthogonal matrices $U_{m\times m}$ and $V_{n\times n}$, as well
as a diagonal one $\Sigma_{m\times n}$ (i.e.\ a non-square matrix such that
$\Sigma_j^{~b}=0$ when $j\neq b$). Consequently,
\be \label{V} 
M M^T = U^T (\Sigma \Sigma^T ) U
\hspace{1cm} \mbox{and} \hspace{1cm}
M^T M = V^T (\Sigma^T \Sigma ) V\; . 
\ee
Therefore, applying $U$ and $V$ respectively on the scalar and gauge boson multiplets
\be \label{uvrot}
\phi_i  = U_{ij} \tilde{\varphi}_j, \hspace{2cm}
W^a_\mu = V^{ab} \tilde{A}^b_\mu
\ee
gives the diagonal mass matrices
\be
m_\phi^2 = \Sigma \Sigma^T =\left[\begin{array}{cc}
D_p &  \\ 
 & 0   \end{array} \right]_{m\times m}
\hspace{1cm} \mbox{and} \hspace{1cm}
m_W^2 = \Sigma^T \Sigma = \left[\begin{array}{cc}
D_p &  \\ 
 & 0 \end{array} \right]_{n\times n}\; .
\ee
Although $m_\phi^2$ and $m_W^2$ are in general of different dimension, this is
only due to their null spaces. The diagonal blocks $D_p$ of dimension
$p=\min(m,n)$ are identical, and include all the non-vanishing entries.

The Lagrangian including the gauge fixing term has now the desired form in the
mass-eigenstate basis:
\bea \label{Lufin}
\Lu_{\rm kin,mass} &=& -\f14 W^T_{\mu\nu} W^{\mu\nu}
                       +\f12 W_\mu^T m_W^2 W^{\mu}
                       +\f12 (\partial_\mu \phi)^T (\partial^\mu \phi)\nnb\\
&&                     -\f{1}{2\xi}(\partial^\mu W_\mu)^T (\partial^\nu W_\nu) 
                       -\f{\xi}{2} \phi^T m_\phi^2 \phi\; .
\eea
The WBG and gauge boson mass matrices are now diagonal. Non-vanishing squared
masses are proportional to each other, with $\xi$ being the proportionality
factor.  The physical scalars are contained in $\phi$ but they do not receive
any mass contribution from the gauge fixing, and therefore correspond to zero
eigenvalues of $m_\phi^2$. As we have already mentioned (below
Eq.~\eqref{psgmass}), contributions to their mass matrix from the full scalar
potential should be added to $\Lu_{\rm kin,mass}$. Obviously, they can be
diagonalized without affecting the r.h.s.\ of Eq.~\eqref{Lufin}.

\section{Ghost sector and BRST} \label{sec:gs}

Our gauge-fixing functionals ${\mathcal G}^a$ in Eq.~\eqref{Lgf} are linear in
the fields. Consequently, the ghost Lagrangian $\Lu_{FP}$ can be derived from
the Fadeev-Popov determinant (see, e.g., section 21.1 of
Ref.~\cite{Peskin:1995ev}). The kinetic terms and interactions for ghosts
$N^a$ and antighosts $\bar N^a$ are then obtained from the variation of
${\mathcal G}^a$ under infinitesimal gauge transformations\footnote{
The gauge couplings in our notation are absorbed into the generators and structure constants.}
$\delta \varphi = -i \alpha^a T^a \left( \varphi + v \right)$~ and~ 
$\delta A_\mu^a = \partial_\mu \alpha^a - f^{abc} A_\mu^b \alpha^c$. 
Taking $\alpha^a(x) = \epsilon N^a(x)$ with an infinitesimal anticommuting
constant $\epsilon$, one gets the BRST~\cite{Becchi:1975nq, Iofa:1976je}
variations
\be \label{sA}
\delta_{\scs \rm BRST} \varphi =  -i \epsilon N^a T^a \left( \varphi + v \right) 
\hspace{1cm} \mbox{and} \hspace{1cm} 
\delta_{\scs \rm BRST} A_\mu^a =  \epsilon \left( \partial_\mu N^a -  f^{abc} A_\mu^b N^c \right)\; .
\ee
The BRST variation of ${\mathcal G}^a$ follows from the above equations, and
can be expressed as
\be \label{sG} 
\delta_{\scs \rm BRST} {\mathcal G}^a = \epsilon M_F^{ab} N^b\; .
\ee
The ghost Lagrangian can now be written in a compact form
\be
\Lu_{FP} = \bar N^a X^{ab} M_F^{bc}\, N^d\label{Lfp}\; ,
\ee
where $X^{ab}$ is an arbitrary field-independent matrix, albeit with a
non-vanishing determinant. A modification of $X^{ab}$ results in
changing the Fadeev-Popov determinant by an irrelevant normalization
constant. For future convenience, we set $X^{ab} = J^{ab}$.  Then our
explicit expression for $\Lu_{FP}$ becomes
\bea
\Lu_{FP} &=&  J^{ab} \bar N^a \raisebox{-.5mm}{$\Box$} N^b 
              + \xi \bar N^a [v^T T^a K T^b v] N^b\nnb\\[1mm]
         &+& \bar N^a\, \mbox{${\raisebox{1.2mm}{\boldmath ${}^\leftarrow$}\hspace{-3mm} \partial}$}{}^{\,\mu}
                         J^{ab} f^{bcd} A_\mu^c N^d
              + \xi \bar N^a [v^T T^a K T^b \varphi ] N^b\; , \label{LuFP}
\eea
where the last two terms describe ghost interactions with the gauge bosons and
scalars.

The BRST variations of ghost and antighosts take the standard form
\be \label{sN}
\delta_{\scs \rm BRST} N^a = \frac{\epsilon}{2} f^{abc} N^b N^c
\hspace{1cm} \mbox{and} \hspace{1cm} 
\delta_{\scs \rm BRST} \bar N^a = \frac{\epsilon}{\xi} {\mathcal G}^a\; . 
\ee
The nilpotence of BRST on $\varphi$ and $A_\mu^a$ follows from Eq.~\eqref{sA}
and $\delta_{\scs \rm BRST} N^a$~\eqref{sN}. A short calculation to check this
fact is exactly the same as in theories without higher-dimensional
operators. Since ${\mathcal G}^a$ is linear in the fields, one concludes that
BRST is nilpotent on ${\mathcal G}^a$, as well, which implies that
$\delta_{\scs \rm BRST} \left( M_F^{ab} N^b \right) = 0$. The latter equality
together with the expression for $\delta_{\scs \rm BRST} \bar N^a$~\eqref{sN}
and Eq.~\eqref{sG} are sufficient to see that
\be
\Lu_{GF} + \Lu_{FP} ~=~ -\f{1}{2\xi}\, {\mathcal G}^a J^{ab} {\mathcal G}^b 
                    ~+~ \bar N^a J^{ab} M_F^{bc} N^c 
\ee
is invariant under BRST, irrespectively of what the actual form of ${\mathcal
G}^a$ is. The remaining parts of the Lagrangian are BRST-invariant thanks to
their gauge invariance.

The matrices parameterizing the ghost kinetic and mass terms in
Eq.~\eqref{LuFP} are identical/proportional to those for the gauge bosons in
Eq.~\eqref{psgmass}. Thus, the ghost mass eigenstates
\be \label{ghosteig}
\eta = V J^\f12 N \hspace{1cm} \mbox{and} \hspace{1cm} \bar \eta =  V J^\f12 \bar N\; .
\ee
are obtained with precisely the same transformations as in Eqs.~\eqref{redef}
and~\eqref{uvrot}, which leads to
\be
\Lu_{FP} ~=~ \bar \eta^T \raisebox{-.5mm}{$\Box$} \eta ~+~ \xi\, \bar \eta^T m_W^2 \eta ~+~ \mbox{(interactions)}\; .
\ee
Ghost masses are thus proportional to the corresponding gauge boson ones, with
$\xi$ being the proportionality factor, as in a theory with no
higher-dimensional operators. However, the ghost interactions and the BRST
variations of antighosts are, in general, affected by the presence of such
operators.

\section{Summary} \label{sec:concl}

We described a procedure for introducing the $R_\xi$ gauge fixing in effective
theories that arise after heavy particle decoupling, taking into account
operators of arbitrarily high dimension. The scalar field VEVs were assumed to
be much smaller than the scale $\Lambda$ whose inverse powers multiply
higher-dimensional terms in the Lagrangian. Treating all such terms as
interactions allowed us to simplify their structure with the help of EOM.  We
showed that it is possible to perform this simplification in such a way that
only operators with single derivatives of the scalar fields $\Phi$, and no
derivatives of the gauge field strength tensor $F_{\mu\nu}^a$ matter for
bilinear terms in $A_\mu^a$ and $\varphi = \Phi - \langle \Phi \rangle$.  They
were parameterized by two matrices depending on $\Phi$ alone, with no
derivatives. Such matrices become constant when $\Phi$ is replaced by its VEV,
and then all the bilinear terms can be resummed into the propagators.\footnote{
Beyond tree level, one renormalizes the two-point one-particle-irreducible
Green's functions treating all the UV counterterms as interactions, including
the EOM-vanishing and/or gauge-variant ones. Next, the renormalized bilinear
terms are the basis for defining the propagators.}

Further steps of our $R_\xi$ gauge-fixing procedure were technically similar
to what one does in theories with initially non-diagonal kinetic terms and
without higher-dimensional operators. Relations between masses of the gauge
bosons, WBG bosons and ghosts remain the same as in the case with canonical
kinetic terms. However, the BRST invariance is maintained only after taking
into account the full dependence on $\Phi$ in the operators that contain the
kinetic and mass terms. Diagonalization of these terms proceeds via field
redefinitions that are not gauge-covariant, and depend on Wilson coefficients
of higher-dimensional operators. For this reason, the ghost terms and BRST
transformations are most conveniently specified before such a diagonalization
is performed. The resulting interactions in the mass-eigenstate basis
(including those of the ghosts) are affected by the presence of
higher-dimensional operators.

\acknowledgments

MM has been partially supported by the National Science Center, Poland, under
the research project 2017/25/B/ST2/00191, and through the HARMONIA project
under contract UMO-2015/18/M/ST2/00518.  MP acknowledges financial support by
the Polish National Science Center under the Beethoven series grant number
DEC-2016/23/G/ST2/04301. JR has been partially supported by the National
Science Center, Poland, under the research projects DEC-2015/19/B/ST2/02848
and DEC-2015/18/M/ST2/00054. KS would like to thank the Greek State
Scholarships Foundation (IKY) for full financial support through the
Operational Programme ``Human Resources Development, Education and Lifelong
Learning, 2014-2020.''

\appendix

\section{The EFT building blocks} \label{app:gi}

In numerous approaches to EFTs with linearly realized gauge symmetries,
higher-dimen-sional operators are constructed from products of gauge field
strength tensors, matter fields, and their covariant derivatives, as dictated
by gauge invariance. However, a frequently asked question is whether any
operator containing non-covariant objects like the usual partial derivatives
could be gauge invariant and, at the same time, not expressible in terms of
covariant derivatives. A very compact (negative) answer to this question was
given in footnote~3 of Ref.~\cite{Grzadkowski:2010es}, while an extended
version can be found in appendix~A of Ref.~\cite{Iskrzynski:2010xxx}. Here, we
recall the relevant argument once again.

If the requirement of gauge invariance was not imposed, a local EFT Lagrangian
density $\Lu$ at a spacetime point $x$ would be a polynomial in fields and
their multiple partial derivatives at this point. For a scalar matter field
$\Phi$, its partial derivative can be trivially re-written in terms of the
covariant one as
\be
\partial_\mu \Phi = \left( \partial_\mu + i A_\mu^a T^a - i A_\mu^a T^a \right) \Phi
                  = D_\mu \Phi - i A_\mu^a T^a \Phi\; .
\ee
Another partial differentiation of this expression gives
\be 
\partial_\nu \partial_\mu \Phi = \left( D_\nu - i A_\nu^b T^b \right) D_\mu \Phi 
- i (\partial_\nu A_\mu^a) T^a \Phi - i A_\mu^a T^a \left( D_\nu - i A_\nu^b T^b \right) \Phi\; ,
\ee
and so on. Thus, $\Lu$ can be re-written in terms of the matter fields, their
covariant derivatives, as well as gauge fields and their multiple partial
derivatives\footnote{
Symmetrization and/or antisymmetrization of $k$ indices in our notation goes with
    a factor of $1/k!\;$.}
\bea
\partial_{ \mu_1} \ldots \partial_{\mu_{k-1}} A^a_{\mu_k } &=& 
\partial_{(\mu_1} \ldots \partial_{\mu_{k-1}} A^a_{\mu_k)} + 
\f{1}{k!} \sum_{\sigma} \left( 
\partial_{ \mu_1} \ldots \partial_{\mu_{k-1}} A^a_{\mu_k } -
\partial_{\mu_{\sigma(1)}} \ldots \partial_{\mu_{\sigma(k-1)}} A^a_{\mu_{\sigma(k)}} \right)\nnb\\
&=& \partial_{(\mu_1} \ldots \partial_{\mu_{k-1}} A^a_{\mu_k)} + 
\f{1}{k} \sum_{j=1}^{k-1} \partial_{\mu_1} \ldots\; \partial_{\mu_j} 
\hspace{-5.5mm}\Big/\hspace{-3mm}\Big\backslash\hspace{3mm} 
\ldots \partial_{\mu_{k-1}} 
\left[ \partial_{\mu_j} A^a_{\mu_k} - \partial_{\mu_k} A^a_{\mu_j} \right]\; .\nnb 
\eea
The last term in the square bracket equals to $F^a_{\mu_j\mu_k} + f^{abc}
A^b_{\mu_j} A^c_{\mu_k}\;$. Under further differentiation, the tensor
$F$ can be treated in the same manner as $\Phi$ above, so only covariant
derivatives of $F$ remain. Thus, further differentiation and subsequent
symmetrization of partial derivatives of $A$ as above will eventually give
us an expression containing $F$ and its covariant derivatives, as well as $A$
and its fully symmetrized partial derivatives only.

At this point, still before imposing gauge invariance on $\Lu$, all the EFT
operators are expressed in terms of matter fields and gauge field strength
tensors, covariant derivatives of them, as well as the fully symmetrized
partial derivatives $\partial_{(\mu_1} \ldots \partial_{\mu_{k-1}}
A^a_{\mu_k)}$, including the zeroth-order one ($k=1$) being equal
to the $A$ field itself.

It remains to be shown that no fully symmetrized derivatives of $A$ can survive
once the gauge-invariance requirement is imposed. One can do this by
considering a series of gauge transformations that sets all such derivatives
($k=1,2,3,\ldots$) to zero at a single but arbitrary spacetime point
$x_P$. We begin with a transformation whose infinitesimal form is
\be 
A^a_\nu(x) \to A^a_\nu(x) + \partial_\nu \alpha^a(x) - f^{abc} A^b_\nu(x) \alpha^c(x) 
\hspace{5mm} \mbox{with} \hspace{5mm} \alpha^a(x) = -(x-x_P)^\rho A^a_\rho(x_P)\; .
\ee
After such a transformation, we have $A^a_\nu(x_P) = 0$. Next, we perform
another transformation choosing $\alpha^a(x) = -\f12 (x-x_P)^\rho
(x-x_P)^\sigma \partial_\rho A^a_\sigma(x_P)$. It preserves the condition
$A^a_\nu(x_P) = 0$ because $\alpha^a(x_P)= \partial_\mu \alpha^a(x_P)
= 0$. Moreover, it nullifies the first symmetrized derivative of $A$ at $x_P$
because $\partial_\mu \partial_\nu \alpha^a = - \partial_{(\mu} A^a_{\nu)}
(x_P)$. Further transformations proceed in an analogous manner. At the $k$-th
step, we choose 
\be
\alpha^a(x) = -\f{1}{k!} (x-x_P)^{\rho_1} \ldots (x-x_P)^{\rho_k}\, \partial_{\rho_1} 
               \ldots \partial_{\rho_{k-1}} A^a_{\rho_k}(x_P)\; . 
\ee
It preserves the conditions $A^a_\nu(x_P) = \partial_{(\mu} A^a_{\nu)}(x_P) =
\ldots = \partial_{(\mu_1}\ldots \partial_{\mu_{k-2}} A^a_{\nu)}(x_P) = 0$
because $\alpha^a(x_P)= \partial_\mu \alpha^a(x_P) = \ldots =
\partial_{\mu_1}\ldots \partial_{\mu_{k-1}} \alpha^a(x_P) = 0$. Moreover, it
nullifies the $(k-1)$-th symmetrized derivative of $A$ at $x_P$ because
$\partial_{\mu_1}\ldots \partial_{\mu_k} \alpha^a = -\partial_{(\mu_1}\ldots
\partial_{\mu_{k-1}} A^a_{\mu_k)}(x_P)$. Working with a full
(non-infinitesimal) form of the gauge transformations would not affect our
arguments because higher-order terms in $\alpha^a$ go with higher powers of
$(x-x_P)$.

We have thus shown that in a particular gauge, any local operator at $x_P$
(even a gauge-variant one) can be written in terms of matter fields, gauge
field strength tensors and their covariant derivatives only. For a gauge
invariant operator, this statement remains true at $x_P$ in any gauge, just
because the operator is gauge invariant by definition. Since the point
$x_P$ was arbitrary, we conclude that gauge invariant local Lagrangian
densities at any point can be written in terms of matter fields, gauge field
strength tensors and their covariant derivatives only.

\section{Distinct gauge-fixing parameters} \label{app:xi}

Our discussion in Sections~\ref{sec:gf}~and~\ref{sec:gs} was restricted to the
case of a single gauge-fixing parameter $\xi$. Here, we generalize it to the
case when distinct gauge-fixing parameters are used for each of the
gauge-boson mass eigenstates. The last two terms of Eq.~\eqref{Lufin} take
then the form
\be \label{distinct}
-\f12(\partial^\mu W_\mu)^T\, \hat{\xi}_D^{-1}\, (\partial^\nu W_\nu) 
-\f12 \phi^T (\Sigma\, \hat{\xi}_D\, \Sigma^T)\, \phi\; , 
\ee
where $\hat{\xi}_D$ is a diagonal matrix with arbitrary but non-vanishing real
entries. Since both $\Sigma_{m\times n}$ and $(\hat{\xi}_D)_{n\times n}$ are
diagonal, it is evident that the scalar mass matrix $(\Sigma\, \hat{\xi}_D\,
\Sigma^T)$ is diagonal. Moreover, all its non-vanishing entries are given by
non-vanishing entries of the diagonal matrix $\;\Sigma^T \Sigma\, \hat{\xi}_D
= m_W^2 \hat{\xi}_D$.

To achieve such a result, we start over with a differently defined $\Lu_{GF}$, namely
\be
\Lu_{GF} = -\f12 {\mathcal G}^a Z^{ab} {\mathcal G}^b\; ,
\ee
with
\be \label{defZ}
Z = J^\f12 V^T \hat{\xi}_D^{-1} V J^\f12 \hspace{1cm} \mbox{and} \hspace{1cm}
{\mathcal G}^a = \partial^\mu A_\mu^a -i (Z^{-1})^{ac} \left[\varphi^T K T^c v\right]\; .  
\ee
The matrix $Z$ is specified in terms the same orthogonal matrix $V$ that
appeared in Eq.~\eqref{svd} for the $\hat{\xi}_D \sim\,${\boldmath $1$} case.

The ``unwanted'' mixing cancels out without making use of the explicit
form of $Z$ in Eq.~\eqref{defZ}, and we arrive at a new version of
Eq.~\eqref{psgmass}, where the only modification is the replacement of $J/\xi$
by $Z$ in the last two terms. Next, the fields get redefined as in
Eq.~\eqref{redef}, which gives us Eq.~\eqref{mmt} with the first three terms
unaltered, and the last two taking the form
\be
-\f12 (\partial^\mu \tilde{A}_\mu)^T V^T \hat{\xi}_D^{-1} V (\partial^\nu \tilde{A}_\nu) 
-\f12 \tilde{\varphi}^T (M V^T \hat{\xi}_D\, V M^T) \tilde{\varphi}\; . 
\ee
Finally, we substitute $M = U^T \Sigma V$, and perform the final rotation of
the fields as in Eq.~\eqref{uvrot}. This way we arrive at
Eq.~\eqref{distinct}.

As far as the ghost terms are concerned, the expression $\Lu_{FP} =
\bar N^a J^{ab} M_F^{bc} N^c$ remains valid. However, $M_F$ defined through
Eq.~\eqref{sG} now depends on $\hat{\xi}_D$ because ${\mathcal G}^a$ in
Eq.~\eqref{defZ} does. Explicitly, one finds
\be
\Lu_{FP} ~=~ \bar N^T J\, \raisebox{-.5mm}{$\Box$} N 
~+~ \bar N^T J^\f12 V^T \hat{\xi}_D V M^T M J^\f12 N ~+~ 
\mbox{(interactions)}\; .
\ee
Diagonalization of the ghost kinetic and mass terms proceeds as in
Eq.~\eqref{ghosteig}, which leads to
\be
\Lu_{FP} ~=~ \bar \eta^T \raisebox{-.5mm}{$\Box$} \eta ~+~ 
             \bar \eta^T m_W^2 \hat{\xi}_D \eta ~+~ \mbox{(interactions)}\; .
\ee
The BRST variations of $\varphi$, $A_\mu^a$ and $N^a$ remain the same as in
Eqs.~\eqref{sA} and \eqref{sN}, while $\delta_{\scs \rm BRST} \bar N^a =
\epsilon \left( J^{-1} Z\right)^{ab} {\mathcal G}^b$.

\section{Scalars in complex representations} \label{app:re}

When setting up our notation in Section~\ref{sec:or}, all the spin-0 degrees
of freedom were expressed in terms of real scalar fields. Such a notation is
not common in the SM and/or SMEFT where scalars furnish complex
representations of the gauge group.  To facilitate re-expressing complex
fields in terms of real ones, we recall a few useful identities below.

For $N$ complex scalar fields denoted collectively by $H$, the
corresponding set of $2N$ real fields $\Phi$ is 
\be
\Phi = {\mathcal U}\Psi, 
\hspace{8mm} \mbox{with} \hspace{8mm}
\Psi = \left( \begin{array}{c} H\\ H^\star \end{array}\right)
\hspace{8mm} \mbox{and} \hspace{8mm}
{\mathcal U} = \frac{S}{\sqrt{2}} \left( \begin{array}{rrr} 
  \mbox{\bf 1}_{\scs N \times N} &&  \mbox{\bf 1}_{\scs N \times N} \\ 
-i\mbox{\bf 1}_{\scs N \times N} && i\mbox{\bf 1}_{\scs N \times N}\end{array}\right)\;, 
\ee
where $S$ is an arbitrary orthogonal $2N\times 2N$ matrix. The matrix
${\mathcal U}$ is unitary.  Denoting the gauge group representation generators
for $H$ by $C^a$, we have $D_\mu H = \left( \partial_\mu + i
A_\mu^a C^a \right)H$. Consequently,~ $D_\mu \Psi = \left(
\partial_\mu + i A_\mu^a P^a \right) \Psi$~ and~ $D_\mu \Phi = {\mathcal
U} D_\mu \Psi = {\mathcal U} \left( \partial_\mu + i A_\mu^a P^a
\right) \Psi = {\mathcal U} \left( \partial_\mu + i A_\mu^a P^a \right)
{\mathcal U}^\dagger \Phi = \left( \partial_\mu + i A_\mu^a T^a \right) \Phi$,
where
\be
P^a = \left( \begin{array}{ccc} C^a &~& \mbox{\bf 0}_{\scs N \times N} \\ 
             \mbox{\bf 0}_{\scs N \times N} && -C^{a\,\star} \end{array}\right) 
\hspace{5mm} \mbox{and} \hspace{5mm} 
T^a ~=~ {\mathcal U} P^a {\mathcal U}^\dagger ~=~ i S \left( \begin{array}{rrr} 
 {\rm Im\,} C^a &~& {\rm Re\,} C^a \\ 
-{\rm Re\,} C^a & & {\rm Im\,} C^a \end{array}\right) S^T\;. 
\ee
Hermiticity of $C^a$ implies that $P^a=P^{a\,\dagger}$ and $T^a =
T^{a\,\dagger}$. Moreover, $T^a$ are manifestly antisymmetric, and all their
components are imaginary.

With the above expressions at hand, any operator containing $H$ and its
covariant derivatives (or their complex conjugates) can easily be expressed in
terms of $\Phi$ and its covariant derivatives. Returning to the notation in
terms of $\Psi$ and then $H$ is also straightforward at any desired
instance.

\section{Gauge fixing in the SMEFT} \label{app:smeft}

As an example, we apply our formalism to the electroweak sector of SMEFT, with
the gauge group $SU(2)\times U(1)$, considered to any fixed order in the
$1/\Lambda$ expansion. Following the notation of Appendix \ref{app:re}, the
complex Higgs doublet and its covariant derivative can be written as
\be
H = \left(\begin{array}{c}
  H^+\\
  H^0
  \end{array}\right)
\equiv
\f{1}{\sqrt{2}}\left(\begin{array}{c}
  \phi_2 + i \phi_1\\
  \phi_4 - i \phi_3
  \end{array}\right)\;,
\hspace{1cm}
D_\mu H = \left(\partial_\mu + \f{i g}{2} \sigma^a W_\mu^a +
\f{i g'}{2} B_\mu \right)H\; .
\ee
For switching to the real notation, we choose
\be
S = \left(\begin{array}{rrrrrrr}
  0 &\;&  0 &\;&  1 &&  0 \\
  1 &&  0 &&  0 &&  0 \\
  0 &&  0 &&  0 && -1 \\
  0 &&  1 &&  0 &&  0
\end{array}\right),
\ee
which gives $\Phi = (\phi_1, \phi_2, \phi_3, \phi_4)$ and
$D_\mu \Phi = \left(\partial_\mu + i T^{a} V^a_\mu\right)\Phi$,
with $V_\mu^a = (W_\mu^1,W_\mu^2,W_\mu^3,B_\mu)$, and
\bea
T^{1} = \f{ig}{2} S \left(\begin{array}{cc} \mbox{\bf 0}_{\scs 2\times 2}  & \sigma^1
  \\ -\sigma^1 & \mbox{\bf 0}_{\scs 2\times 2}  \end{array}\right) S^T\; ,
&\qquad &
T^{2} = \f{g}{2} S \left(\begin{array}{cc} \sigma^2 & \mbox{\bf 0}_{\scs 2\times 2}  \\ 
\mbox{\bf 0}_{\scs 2\times 2}  & \sigma^2 \end{array}\right) S^T\; , \nonumber\\[2mm]
T^{3} = \f{ig}{2} S \left(\begin{array}{cc} \mbox{\bf 0}_{\scs 2\times 2}  & \sigma^3
  \\ -\sigma^3 & \mbox{\bf 0}_{\scs 2\times 2}  \end{array}\right) S^T\; ,
&\qquad &
T^{4} = \f{ig'}{2} S \left(\begin{array}{rrr} 
 \mbox{\bf 0}_{\scs 2\times 2} && \mbox{\bf 1}_{\scs 2\times 2}  \\ 
-\mbox{\bf 1}_{\scs 2\times 2} && \mbox{\bf 0}_{\scs 2\times 2} 
\end{array}\right) S^T\; .
\eea
The matrices $T^a$ are proportional to those in Eq.~(9) of
Ref.~\cite{Helset:2018fgq}. After the Higgs field takes its VEV $\langle \Phi
\rangle = (0,0,0,v)$, the surviving electromagnetic $U(1)_{em}$ gauge
transformations act on the charged gauge bosons as follows:
\bea
W^\pm_\mu = \f{1}{\sqrt{2}} (W^1_\mu \mp i W^2_\mu) \to
e^{\pm i\alpha} \f{1}{\sqrt{2}} (W^1_\mu \mp i W^2_\mu)\; ,
\eea
which is equivalent to 
\bea
\left(\begin{array}{c}
  W^1_\mu \\
  W^2_\mu
\end{array}\right) \to
\left(\begin{array}{rrr}
  \cos\alpha && \sin\alpha \\
  -\sin\alpha && \cos\alpha
\end{array}\right) 
\left(\begin{array}{c}
  W^1_\mu \\
  W^2_\mu
\end{array}\right) \equiv
Q_\alpha \left(\begin{array}{c}
  W^1_\mu \\
  W^2_\mu
\end{array}\right)\; .
\eea
Thus, the gauge boson kinetic matrix $J$ of Eq.~\eqref{eq:gdefc} must be
invariant under the transformation
\bea
\left(\begin{array}{cc}
  Q_\alpha^T & \mbox{\bf 0}_{\scs 2\times 2} \\
  \mbox{\bf 0}_{\scs 2\times 2}  & \mbox{\bf 1}_{\scs 2\times 2} 
\end{array}\right)
J
\left(\begin{array}{cc}
  Q_\alpha & \mbox{\bf 0}_{\scs 2\times 2}  \\
  \mbox{\bf 0}_{\scs 2\times 2}  & \mbox{\bf 1}_{\scs 2\times 2} 
\end{array}\right)
=J\; ,
\eea
which constrains it to the block-diagonal form
\bea
J = \left(\begin{array}{cccc}
   1+J_+ & 0 & 0 & 0 \\
  0 & 1+J_+ & 0 & 0 \\
   0 & 0 & 1+J_1 & J_3 \\
   0 & 0 & J_3 & 1+J_2 \\
\end{array}\right)
\equiv  \left(\begin{array}{cc}
   J_C  & \mbox{\bf 0}_{\scs 2\times 2}\\
  \mbox{\bf 0}_{\scs 2\times 2} & J_N \\
\end{array}\right)\; .
\eea
The same argument ensures identical block-diagonal structure of the scalar
kinetic matrix $K$ and, in consequence, of the matrices $M$,
$U$, $V$ and $\Sigma$ in Eqs.~\eqref{eq:mmatdef} and~\eqref{svd}.

In the charged sector, one finds~ $\Sigma_C = M_W \, \mbox{\bf 1}_{\scs 2\times
2}$~ and~ $U_C = V_C = \mbox{\bf 1}_{\scs 2\times 2}$,~ with the charged
$W$-boson mass squared equal to
\bea
M_W^2 = \f{g^2 v^2}{4}\, \f{1 + K_+}{1 + J_+}\; .
\eea
In this sector, one should use a common gauge parameter
$\xi_W$ to preserve the $U(1)_{em}$ gauge symmetry.

In the neutral sector, let us denote $J'_i = 1 + J_i + \sqrt{{\rm det}
J_N}$, for $i=1,2$. Then one finds
\bea
J_N^{1/2} &=& \f{1}{\sqrt{J'_1 + J'_2}}  \left(\begin{array}{cc} J'_1 &
  J_3 \\ J_3 & J'_2 \\
\end{array}\right)\; ,\nonumber\\
J_N^{-1/2} &=& \f{1}{\sqrt{(J'_1 + J'_2)\, {\rm det} J_N}} \,
\left(\begin{array}{cc} J'_2 & -J_3
  \\ -J_3 & J'_1 \\
\end{array}\right)\; ,
\eea
and similarly for the neutral scalar kinetic matrix $K_N$.  The
matrices appearing in the SVD decomposition~\eqref{svd} for the neutral
sector are:~ $\Sigma_N = {\rm diag}(M_Z,0)$,
\be \label{unvn}
U_N = \left(\begin{array}{rrr}
\cos\omega  && \sin\omega \\
-\sin\omega && \cos\omega \\
\end{array}\right)
\hspace{1cm} \mbox{and} \hspace{1cm}
V_N = 
\left(\begin{array}{rrr}
\cos\theta && -\sin\theta \\
\sin\theta && \cos\theta
\end{array}\right)\; ,
\ee
where~ $\omega = \arctan(K_3/K'_1)$~ and~ $\theta = \arctan[(g' J'_1 + g
J_3)/(g J'_2 + g' J_3)]$.~ In the limit $\Lambda \to \infty$, we have $\omega
\to 0$ and $\theta \to \theta_W \equiv \arctan(g'/g)$. Since ${\mathcal
O}(v/\Lambda)$ effects are small by assumption, both angles should be close to
these limiting values.  The $Z$ boson mass squared equals to
\bea
M_Z^2 = \f{v^2}{4} \left( g^2 + g'{}^2 + g'{}^2 J_1 + 2 g g' J_3 + g^2 J_2 \right) \f{1 + K_1}{{\rm det}\, J_N}\; .
\eea

The above formulae for the SVD matrices and gauge boson masses are valid in
the SMEFT including operators up to any (fixed) dimension. For consistency,
they must always be expanded to the same order in $v/\Lambda$ to which
the matrices $K$ and $J$ are known. Up to ${\mathcal O}(v^2/\Lambda^2)$,
following Ref.~\cite{Dedes:2017zog}, one has:
\bea 
J_+ = J_1 = -\f{2v^2}{\Lambda^2}C^{\varphi W}\; , \qquad
J_2 = -\f{2v^2}{\Lambda^2}C^{\varphi B}\; , \qquad
J_3 = \f{v^2}{\Lambda^2}C^{\varphi WB}\; , 
\eea
\bea
K_+ = K_3 = 0\; , \qquad
K_1 =  \f{v^2}{2\Lambda^2} C^{\varphi D}\; ,\qquad
K_2 = \f{v^2}{2\Lambda^2} (C^{\varphi D} - 4 C^{\varphi\Box })\; .
\eea
After introducing the effective gauge couplings $\bar g = g/\sqrt{1+J_1}$,
$\bar{g}' = g'/\sqrt{1+J_2}$ and expanding in $v/\Lambda$, one recovers the
gauge boson masses, gauge fixing terms and ghost interactions derived in
Ref.~\cite{Dedes:2017zog}.

\end{document}